\documentclass[aps,prl,twocolumn,superscriptaddress,groupedaddress]{revtex4-2}

\usepackage[dvipsnames]{xcolor}
\usepackage{graphicx}
\usepackage{natbib}
\usepackage{amssymb}
\usepackage{amsmath,empheq}
\usepackage{ulem}
\usepackage{dcolumn}
\usepackage{bm}

\newcommand{\bnabla}{\boldsymbol{\nabla}}

\usepackage{cancel}

\begin{document}

\preprint{APS/123-QED}
\title{Equilibrium statistical mechanics of waves in inhomogeneous moving media}%

\author{Alexandre Tlili}%
\author{Basile Gallet}
\affiliation{Université Paris-Saclay, CNRS, CEA, Service de Physique de l’Etat Condensé, 91191 Gif-sur-Yvette, France.}%

\date{\today}

\begin{abstract}


We adapt the microcanonical framework of equilibrium statistical mechanics to predict the statistics of short waves in inhomogeneous moving media. For steady inhomogeneities and background flow, we compute the wave spectrum at any location in the domain based on an ergodic prescription for the action density in phase space, constrained by conservation of absolute frequency. We illustrate the method for shallow-water waves subject to a background flow or to topographic inhomogeneities, and for deep-water surface capillary waves over a background flow, validating the predicted maps of rms surface elevation and interfacial slope against numerical simulations.

\end{abstract}

\keywords{Suggested keywords}

\maketitle



The interplay between waves and mean flows is a fascinating topic with applications ranging from metrology in experimental fluid mechanics -- sound-vorticity or surface-wave-vorticity interaction~\cite{rayleigh1896theory,lund1989ultrasound,cerda1993interaction,berthet1995forward,labbe1998propagation,humbert2017surface,prabhudesai2022statistics} --  to the dynamics of oceanic and atmospheric flows~\cite{vallis2017atmospheric}. 
In the tropical atmosphere, internal gravity waves interacting with the zonal mean-flow induce the quasi-biennial oscillation~\cite{lindzen1968theory,plumb1978instability,baldwin2001quasi,semin2018nonlinear,renaud2019periodicity,leard2020multimodal}.
At the Ocean surface, background mean flows deflect surface waves \cite{kenyon1971wave,gallet2014refraction,ardhuin2017small,Wang2025}, while strong waves feedback onto the mean flow~\cite{craik1976rational,d2014quantifying,seshasayanan2019surface,Buhler2014}. Deeper in the Ocean, the mean flow shapes the inertia-gravity waves arising in the rapidly rotating density stratified fluid~\cite{YBJ1997,balmforth1998enhanced,danioux2015concentration,Kafiabad2019,asselin2020penetration,conn2024interpreting}, with important consequences for vertical mixing~\cite{alford2012annual,jochum2013impact,alford2016near}.  

When the waves are much shorter than the background flow and medium inhomogeneities, ray-tracing is the preferred method to predict wave evolution~\cite{Buhler2014,dong2023geostrophic,Boury2023}, in a similar fashion to geometric optics. When it comes to predicting wave statistics, however, simulating a large-number of rays soon becomes computationally intensive and seems to be missing any overarching large-scale organizing principle. 
In a recent study, we showed that short near-inertial Ocean waves over a background flow behave in an analogous fashion to charged particles in an electromagnetic field, providing a way to predict the wave statistics based on the statistical mechanics of particle systems~\cite{Tlili2025}. For arbitrary waves propagating in an inhomogeneous moving medium, however, no such exact analogy to a particle system is available. The question thus remains whether equilibrium statistical mechanics can be leveraged in this more general context. In the present Letter we answer in the positive, unveiling
an organizing principle based on a parallel with the microcanonical ensemble of statistical mechanics. Namely, combining energy and wave-action conservation for waves in moving media~\cite{Bretherton1968,Buhler2014} with an ergodic hypothesis motivated by chaotic motion in phase space and mathematical results on quantum chaos~\cite{shnirel1974ergodic,voros1976semi,berry1977regular,colin1985ergodicite,zelditch1987uniform,helffer1987ergodicite,kupfer1994ergodic,bouzouina1996equipartition}, we predict the time-averaged statistics of the wave field at any location in space. We illustrate the approach for shallow-water gravity waves and deep-water capillary waves in inhomogeneous or moving media.


\begin{figure*}
    \centering
    \includegraphics[width=\textwidth]{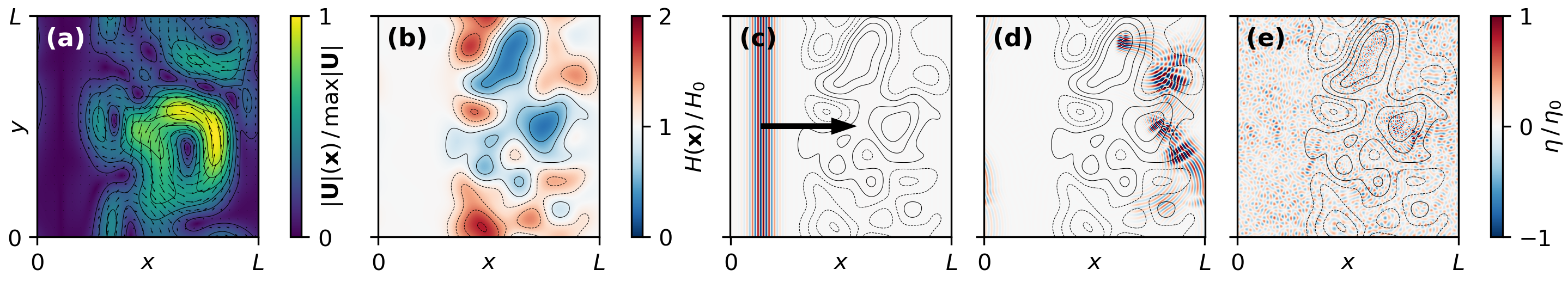}
    \caption{Steady background flow (a) and background topography (b) employed in the various numerical simulations. Panels (c-e) display snapshots of the surface elevation normalized by its maximum initial value $\eta_0$ for a simulation of shallow-water waves above the topography in panel (b).
    The initial rightward-propagating wavepacket (c, $t=0$) gets scattered at early time (d, $t=0.8L/\sqrt{gH_0}$), eventually reaching a statistically steady state at longer time (e, $t=4.9L/\sqrt{gH_0}$). 
    \label{fig:snapshots}}
\end{figure*}

\textit{Micro-canonical statistical mechanics} -- Consider linear waves in an inhomogeneous moving medium. Both the properties of the medium and its velocity are assumed to be time-independent and to vary on a scale much greater than the wavelength. Denoting as $\omega$ and ${\bf k}$ the frequency and wavevector, the wavefield locally satisfies the dispersion relation:
\begin{align}
\omega & = \Omega({\bf x}, {\bf k}) = \hat{\Omega}({\bf x}, {\bf k})  + {\bf U}({\bf x})\cdot {\bf k} \, ,
\end{align}
where ${\bf U}({\bf x})$ denotes the velocity of the background medium. Above, $\omega$ and $\Omega({\bf x},{\bf k})$ are the {\it absolute} frequency and dispersion relation, corresponding to the wave frequency as measured by a steady motionless observer in a fixed reference frame. By contrast, $\hat{\Omega}$ denotes the {\it intrinsic} wave frequency, as measured by an observer moving with the local speed of the background medium. For a motionless medium, $\Omega({\bf x}, {\bf k})=\hat{\Omega}({\bf x}, {\bf k})$, where the ${\bf x}$-dependence of $\hat{\Omega}$ encodes possible spatial inhomogeneities of the properties of the medium.
The scale separation calls for a description of the wave field in terms of wave packets obeying the ray-tracing equations. For a narrow wave-packet located at position ${\bf X}(t)$ with wavevector ${\bf K}(t)$, the latter equations take the Hamiltonian form:
\begin{align}
\dot{\bf X}=\partial_{\bf k} \Omega\, \qquad \dot{\bf K}=- \partial_{\bf x} \Omega \, , \label{eq:ray-tracing}
\end{align}
where the partial derivatives are to be understood component-wise and are evaluated at $({\bf X},{\bf K})$. As a wave packet follows the path given by~(\ref{eq:ray-tracing}), it conserves its wave action, defined as the ratio of the intrinsic wave energy -- energy functional for waves above a motionless background state -- divided by the intrinsic wave frequency \cite{Bretherton1968}. For an ensemble of wave packets with various initial positions and wave numbers, one defines the action density $a({\bf x},{\bf k},t)$ of the ensemble of wave packets in the phase space $({\bf x},{\bf k})$. The Hamiltonian ray-tracing equations~(\ref{eq:ray-tracing}) correspond to a divergence-free flow in  phase space, such that the conservation equation for $a({\bf x},{\bf k},t)$ reduces to the Liouville equation:
\begin{align}
\partial_t a + \partial_{\bf k} \Omega \cdot \partial_{\bf x} a - \partial_{\bf x} \Omega \cdot \partial_{\bf k} a = 0 \, . \label{eq:Liouville}
\end{align}
Because the medium is time-independent, the wave field is governed by partial differential equations with time-independent coefficients. This invariance to time translations results in the conservation of the absolute wave frequency $\omega$ of a given wave packet. We are thus in a position to apply the micro-canonical formulation of equilibrium statistical mechanics, where the action is equivalent to the number of particles, while the frequency plays the role of the energy. Namely, consider an ensemble of wave packets of same absolute frequency $\omega_0$. In phase space, this cloud of wave packets only has access to the hypersurface $\Omega({\bf x}, {\bf k})=\omega_0$ as it gets distorted by the phase-space flow (right-hand side of~(\ref{eq:ray-tracing})). The latter being typically chaotic, it tends to homogenize the action density over the hypersurface. The ergodic hypothesis consists in assuming perfect homogenization, leading to the micro-canonical expression for the time-averaged action density:
\begin{align}
\overline{a}({\bf x},{\bf k})={\cal C} \, \delta \left[  \Omega({\bf x}, {\bf k})  - \omega_0 \right] \, , \label{eq:aergodic}
\end{align}
where the overbar denotes time average and ${\cal C}$ is a normalization factor. 
In the following we consider waves in a doubly periodic domain ${\bf x}\in {\cal D}=[0,L]^2$ with (conserved) total action ${\cal A}$, in which case this normalization factor simply reads ${\cal C}={\cal A}/\int_{{\bf x}\in {\cal D}, {\bf k}\in \mathbb{R}^2}    \delta \left[  \Omega({\bf x}, {\bf k})  - \omega_0 \right]     \mathrm{d}{\bf x} \mathrm{d}{\bf k}$. One easily checks that~(\ref{eq:aergodic}) is a steady solution to the Liouville equation~(\ref{eq:Liouville}). The action density (\ref{eq:aergodic}) provides a way of computing various time-averaged statistics of the wave field through phase-space integration. To illustrate the method, in the following sections we compute the spatial distributions of rms interface displacement and wave slope for various surface-wave systems.

\begin{figure*}
    \centering
    \includegraphics[width=\textwidth]{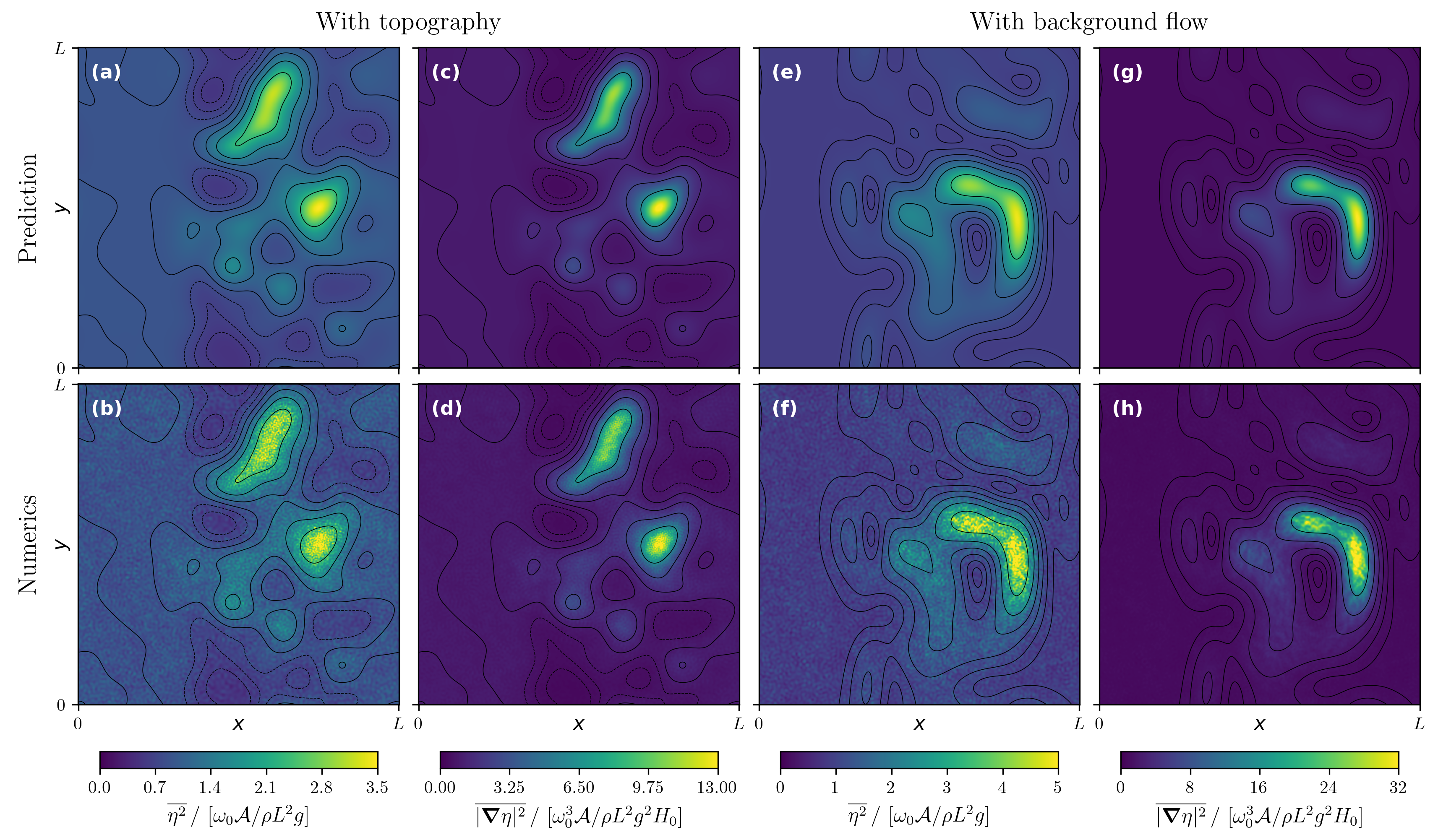}
    \caption{Comparison of the ergodic predictions for $\overline{\eta^2}$ and $\overline{|\bnabla \eta|^2}$ with the numerical simulations for shallow-water waves. Panels (a–d) show the case with topography, while panels (e–h) correspond to the case with a background flow. The top row displays the theoretical predictions \eqref{eq:eta2SW} and \eqref{eq:gradeta2SW}, and the bottom row shows the corresponding numerical simulations. }
    \label{fig:resultsSW}
\end{figure*}

\textit{Application 1: shallow-water waves} -- As a first example, we consider short shallow-water surface waves of frequency $\omega_0$ in a fluid layer of inhomogeneous depth $H(\bf {x})$ hosting a steady background flow ${\bf U}(\bf x)$. The domain is doubly periodic in the horizontal directions, ${\bf x}\in [0,L]^2$. The intrinsic dispersion relation reads~$\hat{\Omega}({\bf x},{\bf k})=\sqrt{g H({\bf x})} k$, where $g$ denotes gravity. The absolute frequency is $\omega_0={\Omega}({\bf x},{\bf k})=\sqrt{g H({\bf x})} k + {\bf U}({\bf x}) \cdot {\bf k}$. Consider waves with wavenumber between ${\bf k}$ and ${\bf k}+\mathrm{d}{\bf k}$ at time $t$ and location ${\bf x}$ in the domain. Denoting the local spectrum of surface elevation as $S_\eta({\bf x}, {\bf k},t)$, the surface elevation variance associated with these waves is $S_\eta({\bf x}, {\bf k},t) \mathrm{d}{\bf k}$, their intrinsic energy density per unit area is $\rho g S_\eta({\bf x}, {\bf k},t) \mathrm{d}{\bf k}$ and the corresponding action per unit area is $a({\bf x},{\bf k},t) \mathrm{d}{\bf k}=\rho g S_\eta({\bf x}, {\bf k},t) \mathrm{d}{\bf k} /(\sqrt{g H({\bf x})} k)$. Time-averaging the latter equality and  substituting the micro-canonical expression~(\ref{eq:aergodic}) for $\overline{a}({\bf x},{\bf k})$ yields the time-averaged surface elevation spectrum:
\begin{align}
\overline{S_\eta}({\bf x}, {\bf k}) = {\cal C} \frac{k \sqrt{H(\bf {x})}}{\rho \sqrt{g}}  \delta \left[  \sqrt{g H({\bf x})} k + {\bf U}({\bf x}) \cdot {\bf k} - \omega_0 \right] \, ,  \label{eq:Seta}
\end{align}
where the expression of the normalization factor ${\cal C}$ is given below equation~(\ref{eq:aergodic}). Integrating over all possible wavenumbers yields the spatial distribution of time-averaged mean-squared surface elevation, $\overline{\eta^2}({\bf x})=\int_{{\bf k}\in \mathbb{R}^2} \overline{S_\eta}({\bf x}, {\bf k})  \, \mathrm{d}{\bf k}$, while the local time-averaged mean-squared interfacial slope is given by $\overline{|\bnabla \eta|^2}({\bf x})=\int_{{\bf k}\in \mathbb{R}^2} k^2 \overline{S_\eta}({\bf x}, {\bf k})  \, \mathrm{d}{\bf k}$. The integrals are computed using polar coordinates in ${\bf k}$ space in the Supplementary Information. 
Denoting as $\tilde{U}({\bf x})=|{\bf U}({\bf x})|/\sqrt{g H({\bf x})}$ the ratio of the local background flow speed to the local wave velocity, the end results read
\begin{align}
\overline{\eta^2}({\bf x}) & = \frac{{\cal A} \tilde{\cal C} \omega_0}{\rho g H({\bf x})} \times \frac{1+\frac{1}{2}\tilde{{ U}}^2({\bf x})}{\left[1-\tilde{{ U}}^2({\bf x}) \right]^{5/2}}\, ,  \label{eq:eta2SW} \\
\overline{|\bnabla \eta|^2}({\bf x}) & = \frac{{\cal A} \tilde{\cal C} \omega_0^3}{\rho g^2 H^2({\bf x})} \times \frac{1+3 \tilde{{ U}}^2({\bf x})+\frac{3}{8}\tilde{{ U}}^4({\bf x})}{\left[1-\tilde{{ U}}^2({\bf x}) \right]^{9/2}} \, ,  \label{eq:gradeta2SW}
\end{align}
where $\tilde{\cal C} = 1/  \int_{\cal D} H^{-1}({\bf x})  [1-\tilde{{ U}}^2({\bf x})]^{-3/2} \mathrm{d}{\bf x}$.

To test these predictions, we performed numerical simulations of the shallow-water equations linearized around a steady background state in a doubly periodic domain $[0,L]^2$ using a standard pseudo-spectral solver (see Supplementary Information for detailed numerical methods).

In a first simulation we consider topographic inhomogeneities around a mean value $H_0$, and a motionless background state ${\bf U}=\boldsymbol{0}$. The topography $H({\bf x})$ is generated from Fourier modes with random phases and amplitudes proportional to $e^{-(kL_{\text{corr}})^2/4}$ with $L_{\text{corr}}=L/2$. It is scaled so that the minimum layer depth is $0.26H_0$.
In a second simulation we consider a background flow only and uniform layer depth $H({\bf x})=H_0$. The flow ${\bf U}({\bf x})$ stems from a streamfunction generated following the same procedure as the topographic inhomogeneities of the first simulation, using $L_\text{corr}=L$. The maximum background flow velocity is set to $\max_{{\bf x}\in {\cal D}}|{\bf U}({\bf x})|/\sqrt{g H_0}=0.67$.
Additionally, for both simulations we taper the background state such that it is  inhomogeneous in the right-hand part of the domain only, while it is uniform over the leftmost region of the domain (see figure~\ref{fig:snapshots}a,b). In the latter region we initialize the wavefield with a rightward-propagating narrow wavepacket of frequency $\omega_0= 40 \sqrt{g H_0}/L$, which corresponds to an initial wavelength much smaller than the scale of the background flow (see figure~\ref{fig:snapshots}c).

The waves scatter as they impinge on the inhomogeneities of the background state, be they flow structures or topographic features (figure \ref{fig:snapshots}b), and the wavefield reaches a statistically steady state in the long-time limit (figure \ref{fig:snapshots}c). 
In figure \ref{fig:resultsSW} we report maps of the mean-squared interface displacement $\overline{\eta^2}({\bf x})$ and mean-squared interfacial slope $\overline{|\bnabla \eta|^2}({\bf x})$, obtained through time-averaging in the statistically steady state of the simulations. These maps agree closely with the theoretical predictions~(\ref{eq:eta2SW}) and~(\ref{eq:gradeta2SW}).

\begin{figure*}
    \centering
    \includegraphics[width=\textwidth]{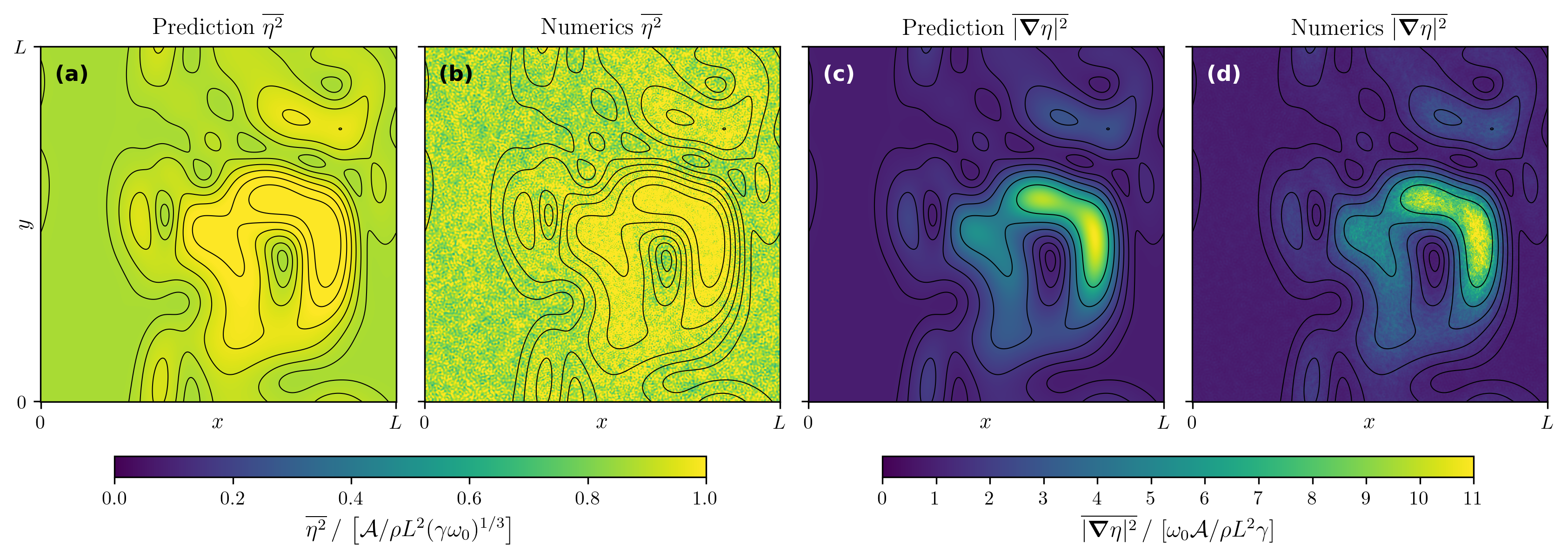}
    \caption{Comparison of the ergodic predictions with numerical simulations for capillary waves over a background flow ${\bf U}({\bf x})$. Panels (a,b) show the predicted and observed $\overline{\eta^2}$, respectively, while panels (c,d) show the predicted and observed $\overline{|\bnabla \eta|^2}$.}
    \label{fig:resultsCapillary}
\end{figure*}

\textit{Application 2: deep-water surface capillary waves} -- With the goal of illustrating the method on fully dispersive waves, we now turn to short deep-water surface capillary waves of absolute frequency $\omega_0$ propagating above a steady large-scale background flow ${\bf U}(\bf x)$. Denoting surface tension as $\sigma$, the intrinsic dispersion relation reads $\hat{\Omega}({\bf k})=\sqrt{\gamma k^3}$, where $\gamma=\sigma/\rho$. The absolute frequency is $\omega_0={\Omega}({\bf x},{\bf k})=\sqrt{\gamma k^3} + {\bf U}({\bf x}) \cdot {\bf k}$. At location ${\bf x}$ and time $t$, the intrinsic energy per unit area of the waves whose wavenumber is between ${\bf k}$ and ${\bf k}+\mathrm{d}{\bf k}$ is $\sigma k^2 S_\eta({\bf x},{\bf k},t) \mathrm{d} {\bf k}$. Dividing by the intrinsic frequency leads to the action per unit area $a({\bf x},{\bf k},t) \mathrm{d}{\bf k}=\rho\sqrt{ \gamma  k} S_\eta({\bf x},{\bf k},t) \mathrm{d} {\bf k}$. Time-averaging and substituting the micro-canonical prescription~(\ref{eq:aergodic}) yields the time-averaged surface elevation spectrum at position ${\bf x}$:
\begin{align}
\overline{S_\eta}({\bf x}, {\bf k}) =  \frac{{\cal C}}{\rho\sqrt{\gamma k}}  \delta \left[  \sqrt{\gamma k^3}  + {\bf U}({\bf x}) \cdot {\bf k} - \omega_0 \right]  \, .  \label{eq:Setacap}
\end{align}
The mean-squared interfacial displacement and interfacial slope are obtained as $\overline{\eta^2}({\bf x})=\int_{{\bf k}\in \mathbb{R}^2} \overline{S_\eta}({\bf x}, {\bf k})  \, \mathrm{d}{\bf k}$ and $\overline{|\bnabla \eta|^2}({\bf x})=\int_{{\bf k}\in \mathbb{R}^2} k^2 \overline{S_\eta}({\bf x}, {\bf k})  \, \mathrm{d}{\bf k}$, respectively. As detailed in the Supplementary Information, denoting the dimensionless speed of the background flow as $\tilde{U}({\bf x})=|{\bf U}({\bf x})|/(\gamma \omega_0)^{1/3}$ these integrals are evaluated as:
\begin{align}
\overline{\eta^2}({\bf x}) & = \frac{{\cal A}\tilde{\cal C}}{\rho(\gamma \omega_0)^{1/3}} \times {\cal F}_{1/2}(\tilde{U}({\bf x})) \, ,  \label{eq:eta2cap} \\
\overline{|\bnabla \eta|^2}({\bf x}) & = \frac{{\cal A}\tilde{\cal C} \omega_0}{\rho \gamma} \times {\cal F}_{5/2}(\tilde{U}({\bf x})) \, ,  \label{eq:gradeta2cap} \\
\text{with } {\cal F}_{\alpha}(X) & = 4 \int_{s_\text{min}}^{s_\text{max}} \frac{s^{2\alpha+1}}{\sqrt{X^2 s^4-(1-s^3)^2}} \mathrm{d}s \, ,
\label{eq:cap_Falpha}
\end{align}
where $[s_\text{min}, s_\text{max}]$ denotes the interval of $\mathbb{R}^+$ over which the radicand is positive and we have introduced the constant $\tilde{\cal C}=1/\int_{\cal D} {\cal F}_{1}(\tilde{U}({\bf y})) \mathrm{d}{\bf y}$.

These predictions call for validation based on laboratory data or fully 3D numerical models. As a first step in this direction, we design and numerically solve a set of reduced 2D equations describing short linear capillary waves over a large-scale background flow. 
To wit, we denote as $\phi(x,y,t)$ the velocity potential evaluated at the free surface and we introduce the operator $D^\alpha$ corresponding to multiplication by $k^{\alpha}$ in spectral space. Hamiltonian theory~\cite{pushkarev2000turbulence,Nazarenko2011} points to the complex variable $\psi(x,y,t) = \gamma ^{1/4}D^{1/4}\eta + i \gamma^{-1/4}D^{-1/4}\phi$, related to the total action through ${\cal A}=\frac{1}{2}\rho \int_D |\psi|^2 \mathrm{d}{\bf x}$. In the absence of background flow, through an inverse Fourier transform of the intrinsic dispersion relation one recovers the standard evolution equation $\partial_t \psi = -i\gamma^{1/2}D^{3/2}\psi$ governing the linear dynamics of $\psi$. 
In the presence of a background flow, an inverse Fourier transform of the Doppler-shifted (absolute) dispersion relation points to the evolution equation $\partial_t \psi + {\bf U}\cdot \bnabla \psi = -i\gamma^{1/2}D^{3/2}\psi$, which conserves the total wave action ${\cal A}$.
We solve the latter equation numerically using the same background-flow geometry as for the shallow-water case (see figure \ref{fig:snapshots}b) and a similar wave-packet initial condition. The frequency is $\omega_0=476 \sqrt{\gamma / L^3}$ and the maximum background flow speed is $\max_{\cal D} \tilde{U}({\bf x})=2.0$. Once the transient has subsided, we extract the time-averaged spatial distributions of mean-squared interface displacement and interfacial slope in the statistically steady state. In figure~\ref{fig:resultsCapillary}, we compare the resulting spatial distributions to the predictions~(\ref{eq:eta2cap}-\ref{eq:gradeta2cap}) of equilibrium statistical mechanics, obtaining again very good agreement.

\textit{Conclusion} -- We have adapted the microcanonical framework of equilibrium statistical mechanics to waves in inhomogeneous moving media, before demonstrating the predictive skill of this approach for two idealized surface-wave systems. For simplicity, we have considered waves at a single frequency, but the approach easily carries over to a superposition of waves at various frequencies. As compared to standard approaches for predicting wave statistics in inhomogeneous media \cite{ryzhik1996transport,Danioux2016,Kafiabad2019,boas2020directional,kafiabad2021interaction,onuki2024dynamical}, an appealing aspect of the present method is that it does not require averaging over an ensemble of realizations of the disorder. Instead, we obtain the wave statistics for a given realization of the background flow. This could be of interest in geophysical settings, such as waves in Ocean currents, where the background flow often evolves over a much slower timescale than the wave field.

In the shallow-water example we have restricted attention to mean flows that are slower than the wave group velocity. This appears necessary to keep finite integrals. Indeed, for wave systems with an intrinsic dispersion relation shallower than (or equal to) $\hat{\Omega} \sim k$, above a threshold speed of the background flow wave action can escape to arbitrarily large wavenumber $k$, in a similar fashion to the `slow-fast transition' recently unveiled in wave systems using numerical ray-tracing \cite{Boury2023}. The present approach could therefore shed new light on this transition. Of crucial interest would also be the inclusion of nonlinearity back into the system \cite{Michel2017,Vernet2025}. Guided by wave turbulence theory \cite{zakharov1992,Nazarenko2011,newell2011wave,FalconMordant2022,galtier2022physics}, we expect action conservation for systems with four-wave interactions (e.g. surface gravity waves), while we expect slow evolution of the action for weakly nonlinear systems governed by three-wave interactions (e.g. surface capillary waves). Finally, beyond wave-wave interactions nonlinear processes potentially include feedbacks of the wavefield onto the background flow \cite{Xie2015,Buhler2014,andrews1978exact,wagner2016three,rocha2018stimulated,wagner2021near}, opening up additional natural directions for future research.

\bibliographystyle{apsrev4-2}
\bibliography{phyStat}

\end{document}


\maketitle

We provide additional technical details on the derivation of the ergodic predictions and on the numerical methods employed to validate them. 


\paragraph{Integration in phase space: shallow-water waves.} Once the time-averaged surface elevation spectrum $\overline{S_{\eta}}(\bx, \bk)$ is related to the microcanonical distribution of wave-action density in phase-space $(\bx, \bk)$, 
one is led to compute integrals of the form $\int_{\mathbb{R}^2} \mathrm{d}^2\bk \, \, k^{\alpha} \delta \left[  \sqrt{gH(\bx)}k  + {\bf U}({\bf x}) \cdot {\bf k} - \omega_0 \right]$. Equations \eqref{eq:eta2SW} and \eqref{eq:gradeta2SW} from the article correspond to $\alpha = 1$ and $\alpha=3$, respectively, while the normalization coefficient $\mathcal{C}$ is evaluated using $\alpha=0$. Assuming $\omega_0 > 0$ and introducing a local dimensionless wavenumber $\tilde{\bk}=\bk \sqrt{gH(\bx)}/\omega_0$ leads to
\begin{align*}
    \int_{\mathbb{R}^2} \mathrm{d}^2\bk \,\, k^{\alpha} \delta \left[  \sqrt{gH(\bx)}k + {\bf U}({\bf x}) \cdot {\bf k} - \omega_0 \right] 
    &= \left( \frac{\omega_0}{\sqrt{gH(\bx)}} \right)^{\alpha+2} \int_{\mathbb{R}^2} \mathrm{d}^2\tilde{\bk} \,\, \tilde{k}^{\alpha} \delta \left[  \omega_0\tilde{k} + \omega_0 \frac{{\bf U}({\bf x})}{\sqrt{gH(\bx)}} \cdot \tilde{{\bf k}} - \omega_0 \right] \\
    &= \frac{1}{\omega_0} \left( \frac{\omega_0}{\sqrt{gH(\bx)}} \right)^{\alpha+2} \int_{\mathbb{R}^2} \mathrm{d}^2\tilde{\bk} \,\, \tilde{k}^{\alpha} \delta \left[  \tilde{k} + \tilde{\bU}(\bx) \cdot \tilde{{\bf k}} - 1 \right] \, ,
\end{align*}
where $\tilde{\bU}(\bx) = \bU(\bx)/\sqrt{gH(\bx)}$. The position $\bx$ being fixed, we introduce local polar coordinates in ${\bf k}$-space, denoting as $\theta$ the angle between $\tilde{\bk}$ and $\tilde{\bU}(\bx)$. The argument of the $\delta$ distribution becomes $\tilde{k} [ 1 + |\tilde{\bU}(\bx)|\cos \theta] - 1$, which vanishes for $\tilde{k} = 1/( 1 + |\tilde{\bU}(\bx)|\cos \theta) > 0$ provided $|\tilde{\bU}(\bx)|<1$. The dimensionless integral can be computed as
\begin{align*}
     \int_{\mathbb{R}^2} \mathrm{d}^2\tilde{\bk} \,\, \tilde{k}^{\alpha} \delta \left[  \tilde{k} + \tilde{\bU}(\bx) \cdot \tilde{{\bf k}} - 1 \right] 
     &= \int_{-\pi}^{\pi} \mathrm{d}\theta \int_0^{\infty} \mathrm{d}\tilde{k} \, \, \tilde{k}^{\alpha+1} \delta \left[ \tilde{k} ( 1 + |\tilde{\bU}(\bx)|\cos \theta ) - 1 \right] \\
     &= \int_{-\pi}^{\pi} \frac{\mathrm{d}\theta}{\left( 1 + |\tilde{\bU}(\bx)|\cos \theta  \right)^{\alpha+2}}  \triangleq f_{\alpha}(|\tilde{\bU}(\bx)|) \, .
\end{align*}
This integral is finite for $|\tilde{\bU}(\bx)|<1$. It can be evaluated exactly for $\alpha=0, 1, 3$, the result being:
\begin{equation*}
    f_{0}(\tilde{U}) = \frac{2\pi}{(1-\tilde{U}^2)^{3/2}}, \quad 
    f_{1}(\tilde{U}) = \frac{2\pi\left(1+\frac{1}{2}\tilde{U}^2\right)}{(1-\tilde{U}^2)^{5/2}}, \quad 
    f_{3}(\tilde{U}) = \frac{2\pi \left( 1 + 3\tilde{U}^2 + \frac{3}{8} \tilde{U}^4 \right)}{(1-\tilde{U}^2)^{9/2}} \, .
\end{equation*}
Substitution of these expressions leads to the value of the constant $\tilde{\mathcal{C}}$ provided in the article, together with the predictions (\ref{eq:eta2SW}-\ref{eq:gradeta2SW}).

\paragraph{Integration in phase space: capillary waves.} Following a similar approach for capillary waves leads to integrals of the form 
$\int_{\mathbb{R}^2} \mathrm{d}^2\bk \,\, k^{\alpha-1} \delta \left[  \sqrt{\gamma k^3}  + {\bf U}({\bf x}) \cdot {\bf k} - \omega_0 \right]$.  Equations~\eqref{eq:eta2cap}, \eqref{eq:gradeta2cap} and the expression for the normalization constant $\tilde{\mathcal{C}}$ in the article correspond to $\alpha=1/2, \, 5/2, \,1 $, respectively. Introducing a dimensionless wavenumber $\tilde{\bk}={\bk} (\omega_0^2/\gamma)^{-1/3}$ leads to
\begin{align*}
    \int_{\mathbb{R}^2} \mathrm{d}^2\bk \,\, k^{\alpha-1} \delta \left[  \sqrt{\gamma k^3}  + {\bf U}({\bf x}) \cdot {\bf k} - \omega_0 \right] = \frac{1}{\omega_0}\left( \frac{\omega_0^2}{\gamma} \right)^{(\alpha+1)/3} \int_{\mathbb{R}^2} \mathrm{d}^2\tilde{\bk} \,\, \tilde{k}^{\alpha-1} \delta \left[ \tilde{k}^{3/2}  + \tilde{\bU}(\bx) \cdot \tilde{\bf k} - 1 \right] \, ,
\end{align*}
where we defined $\tilde{\bU}(\bx) = {\bf U}({\bf x})/(\gamma \omega_0)^{1/3}$. When $\tilde{\bU}={\bf 0}$ the value of the integral is $4\pi/3$. When $\tilde{\bU} \neq {\bf 0}$, progress can be made by using a local coordinate system aligned with $\tilde{\bU}(\bx)$. Denoting as $\tilde{k}_u$ the streamwise component of $\tilde{\bk}$, we have $\tilde{\bU}(\bx) \cdot \tilde{\bk} = |\tilde{\bU}(\bx)|\tilde{k}_u$. Using $(\tilde{k},\tilde{k}_u)$ as integration variables, we get
\begin{align*}
    \int_{\mathbb{R}^2} \mathrm{d}^2\tilde{\bk} \,\, \tilde{k}^{\alpha-1} \delta \left[ \tilde{k}^{3/2}  + \tilde{\bU}(\bx) \cdot \tilde{\bf k} - 1 \right] 
    &= 2\int_0^{\infty} \mathrm{d}\tilde{k} \int_{-\tilde{k}}^{\tilde{k}} \mathrm{d}\tilde{k}_u \,\, \frac{\tilde{k}^\alpha}{\sqrt{\tilde{k}^2 - \tilde{k}_u^2}} \delta \left[ |\tilde{\bU}(\bx)| \tilde{k}_u - (1 - \tilde{k}^{3/2}) \right] \, ,
\end{align*}
where the factor $\tilde{k}/\sqrt{\tilde{k}^2 - \tilde{k}_u^2}$ is the Jacobian associated with the change of variables. For each value of $\tilde{k}>0$, the inner integral is non-zero if and only if $(1 - \tilde{k}^{3/2})^2 < \tilde{k}^2\tilde{U}^2(\bx)$, which is satisfied on an open interval $\left( \tilde{k}_\text{min}(\tilde{U}), \, \tilde{k}_\text{max}(\tilde{U}) \right)$ only. A final change of variable to $s=\tilde{k}^{1/2}$ yields:
\begin{equation*}
    \int_{\mathbb{R}^2} \mathrm{d}^2\tilde{\bk} \,\, \tilde{k}^{\alpha-1} \delta \left[ \tilde{k}^{3/2}  + \tilde{\bU}(\bx) \cdot \tilde{\bf k} - 1 \right] = 2\int_{\tilde{k}_\text{min}}^{\tilde{k}_\text{max}}  \frac{\tilde{k}^\alpha}{\sqrt{\tilde{U}(\bx)^2\tilde{k}^2 - (1 - \tilde{k}^{3/2})^2}} \mathrm{d}\tilde{k} = 4\int_{s_\text{min}}^{s_\text{max}}  \frac{s^{2\alpha+1} \, \mathrm{d}s}{\sqrt{\tilde{U}(\bx)^2 s^4 - (1 - s^3)^2}} \, .
\end{equation*}
where $s_{\min} = \tilde{k}_{\min}^{1/2}$ and $s_{\max} = \tilde{k}_{\max}^{1/2}$. The right-hand side is precisely the function $\mathcal{F}_{\alpha}$ defined in Equation~\eqref{eq:cap_Falpha} of the article, evaluated at $X=|\tilde{\bU}(\bx)|$. The bounds $s_\text{min}(\tilde{U})$ and $s_\text{max}(\tilde{U})$ are the only positive roots of the polynomials $s^3+\tilde{U}s^2 - 1$ and $s^3-\tilde{U}s^2 - 1$, respectively. They can be computed exactly using Cardans's formula.

\paragraph{Numerical setup: shallow-water waves.} The dimensional  shallow-water equations linearized around a background state with velocity $\bU(\bx)$ and total layer thickness $H(\bx)$ (including variations in both topography and surface elevation) read
\begin{equation*}
    \left\{
    \begin{aligned}
        &\partial_t \eta + \bnabla \cdot (\eta \bU) + \bnabla \cdot (H \bu) = 0 \, , \\
        &\partial_t \bu + (\bU \cdot \bnabla) \bu + (\bu \cdot \bnabla) \bU = - g\bnabla \eta \, ,
    \end{aligned}
    \right.
\end{equation*}
where $\eta(\bx,t)$ and $\bu(\bx,t)$ respectively denote the water height perturbation and the velocity perturbation associated with the waves. We non-dimensionalize $\eta$ using the maximum surface displacement $\eta_0$ in the initial state. Accordingly, we non-dimensionalize the wavy flow $\bu$ using $\eta_0\sqrt{g/H_0}$.
Denoting the domain size as $L$, we non-dimensionalize $\bU(\bx)$, $H(\bx)$, $t$ and $\bx$ using $\sqrt{gH_0}$, $H_0$, $L/\sqrt{gH_0}$ and $L$, respectively. Replacing variables with their dimensionless counterparts for brevity 
the dimensionless equations read:
\begin{equation*}
    \left\{
    \begin{aligned}
        \partial_t \eta + \bnabla \cdot (\eta \bU) + \bnabla \cdot (H \bu) &= 0 \, , \\
        \partial_t \bu + (\bU \cdot \bnabla) \bu + (\bu \cdot \bnabla) \bU &= -\bnabla \eta \, .
    \end{aligned}
    \right.
\end{equation*}
We solve these equations in a doubly-periodic domain using the methods detailed below.

\paragraph{Numerical setup: capillary waves.} We numerically simulate the evolution of short capillary waves over a larger-scale steady background flow $\bU(\bx)$. Denoting the surface elevation and the surface velocity potential as $\eta(x,y,t)$ and $\phi(x,y,t)$, respectively, we evolve the complex field $\psi(x,y,t) = \gamma^{1/4}D^{1/4}\eta + i\gamma^{-1/4}D^{-1/4}\phi$ according to $\partial_t \psi + {\bf U} \cdot \bnabla \psi = -i\gamma^{1/2}D^{3/2}\psi$. We non-dimensionalize space, time, and velocities using the characteristic scales $L$, $(L^3/\gamma)^{1/2}$, and $(\gamma/L)^{1/2}$, respectively. The maximum value $\eta_0$ of the initial wavy surface displacement sets the scales $\eta_0$, $\eta_0 (\gamma/L)^{1/2}$ and $\eta_0 (\gamma/L)^{1/4}$ employed to non-dimensionalize $\eta$, $\phi$ and $\psi$, respectively. This procedure leads to the dimensionless evolution equation $\partial_t \psi + {\bf U} \cdot \bnabla \psi = -iD^{3/2}\psi$ for the dimensionless complex field $\psi(x,y,t) = D^{1/4}\eta + iD^{-1/4}\phi$. We solve this governing equation in a doubly-periodic domain using the numerical methods detailed below.

\paragraph{Numerical methods.} We employ a standard pseudo-spectral method, evaluating  spatial derivatives in spectral space and products in real space. Aliasing errors are removed using the standard $2/3$ rule. Time integration is carried out with a classical explicit fourth-order Runge--Kutta scheme. The spatial resolution ranges from $1024^2$ to $2048^2$ collocation points, depending on the run (see Table~\ref{tab:sims}).

All Fast Fourier Transforms and pointwise operations are parallelized on GPUs to achieve efficient performance at high resolution. Time averages of quadratic diagnostic quantities such as $\overline{\eta^2}$ and $\overline{|\bnabla \eta|^2}$ are integrated on the fly during the numerical run and saved at regular intervals. No hyper-viscosity is required; the simulations are numerically stable without it, the spectra of the various fields displaying an emergent cutoff wavenumber well-within the resolved range. A summary of the parameter values employed for the simulations reported in the article is provided in Table~\ref{tab:sims}.

\begin{table}[]
    \centering
    \begin{tabular}{|l|c|c|c|c|c|}
    \hline
    Description & $N$ & $dt \, / \, t_{*}$ & $t_{\min} \, / \, t_{*}$ & $t_{\max} \, / \, t_{*}$ & $t_{\max} \, / \, t_{\text{cross}}$\\
    \hline
    Shallow water waves with topography & $1024^2$ & $1.6\times10^{-5}$ & $4.8$ & $8.0$ & $8.0$ \\
    \hline
    Shallow water waves above a steady flow & $2048^2$ & $1.6\times10^{-5}$ & $4.8$ & $9.5$ & $9.5$ \\
    \hline
    Capillary waves above a steady flow & $1024^2$ & $6.3\times10^{-7}$ & $0.32$ & $0.63$ & $18.6$ \\
    \hline
    \end{tabular}
    \caption{\textbf{Simulation parameters.} $N$ denotes the number of Fourier modes, $dt$ the time step, and $[t_{\min}, \, t_{\max}]$ the dimensional time interval used to compute the time-averaged fields $\overline{\eta^2}$ and $\overline{|\bnabla \eta|^2}$. 
    The variable $t_{*}$ denotes the reference time used for non-dimensionalization of the governing equations. By contrast, the characteristic time $t_{\text{cross}}$ denotes the time it would take the initial wave packet to travel over a distance $L$, were the medium steady and homogeneous. For shallow-water waves, $t_{*} = t_{\text{cross}} =L/\sqrt{gH_0}$. For capillary waves, $t_{*} = \sqrt{L^3/\gamma}$ and $t_{\text{cross}}= \frac{2}{3} L/(\gamma\omega_0)^{1/3}$.
    }
    \label{tab:sims}
\end{table}


\bibliographystyle{apsrev4-2}